\documentclass[runningheads,a4paper]{llncs}

\usepackage[T1]{fontenc}
\usepackage{amssymb}
\usepackage{graphicx}   

\usepackage{url}    
\urldef{\mailsa}\path|{anna.bernasconi,claudio.menghi,carlo.ghezzi}@polimi.it|
 \urldef{\mailsc}\path|lenore@cs.uic.edu|   
\urldef{\mailsd}\path|lmenghi@chalmers.se|

\usepackage{mathpartir}
\usepackage[inline]{enumitem}
\usepackage{cancel}
\usepackage{algorithm}
\usepackage{longtable}
\usepackage{algpseudocode}
\usepackage{todonotes}
 \usepackage{ltl}
\usepackage{multicol}
\usepackage{amsmath}
\usepackage{textgreek}
\usepackage{enumitem}
\usepackage{pifont}
\usepackage{changebar}
\usepackage{changes}
\usepackage{multirow}
\usepackage{makecell}

\definecolor{mygreen}{rgb}{0.0, 0.5, 0.0}
\definecolor{myred}{rgb}{0.8, 0.0, 0.0}

\usepackage{scalerel}

\author{Claudio Menghi}
\author{Anna Bernasconi}

\usetikzlibrary{automata,positioning}
\usetikzlibrary{shapes,shapes.geometric}
\usetikzlibrary{backgrounds}

\tikzset{state/.style={
        draw, 
        ellipse, 
        minimum height=2em, 
        minimum width=2.8em,  
        thick } }
\tikzset{statep/.style={
    draw, 
    ellipse, 
    minimum height=1em, 
    minimum width=1em,  
    thick } }
\tikzset{emptynode/.style={ draw } }
\tikzset{empty/.style={ } } 

\usepackage[normalem]{ulem}

\newcommand{\NAME}{THRIVE}

\usepackage[skip=0pt]{caption}
\begin{document}
\captionsetup[longtable]{labelfont=bf}
\captionsetup[table]{labelfont=bf}
\captionsetup[figure]{labelfont=bf}

\title{From model checking to a temporal proof \\ for partial models: preliminary example}

\titlerunning{From model checking to a temporal proof for partial models}
\authorrunning{From model checking to a temporal proof for partial models}

%
%

\author{Anna Bernasconi\inst{1}
\and Claudio Menghi\inst{2}
\and Paola Spoletini\inst{3}
\and \\ Lenore D. Zuck\inst{4}
\and Carlo Ghezzi\inst{1}
}

\institute{
Politecnico di Milano, DEIB - DEEPSE group,\\
\email{\{anna.bernasconi, carlo.ghezzi\}@polimi.it}\\
\and
Chalmers University of Technology | University of Gothenburg\\
\email{claudio.menghi@gu.se}
\and
Kennesaw State University,\\
\email{pspoleti@kennesaw.edu}\\
\and
University of Illinois at Chicago,\\
\email{lenore@cs.uic.edu}\\
}

\maketitle

\begin{abstract}
This paper describes in detail the example introduced in the preliminary evaluation of \NAME.
Specifically, it evaluates \NAME\ over an abstraction of the ground model proposed for a critical component belonging to a medical device used by optometrists and ophtalmologits to dected visual problems. 
\end{abstract}

We provide the full description of the example introduced in the preliminary evaluation of~\cite{Bernasconi}.
Specifically, we evaluate \NAME\ over an abstraction of the ground model proposed in \cite{arcaini2015formal}, a critical component belonging to a medical device used by optometrists and ophtalmologits to dected visual problems. 
In the following we describe the considered partial model, the property of interest and the deductive verification procedure performed by \NAME\ over the incomplete model.

\vskip 0.05in  
\textbf{Partial model.} The ground model proposed in \cite{arcaini2015formal} is a critical component that measures the stereoacuity of young patients. The criticality of the system resides in certifying a certain level of stereoacuity in a consistent way, such that the treatment given by the doctor to his/her patient is correct. 

\begin{figure}[h!]
\begin{center}         
 	\includegraphics[width=\linewidth]{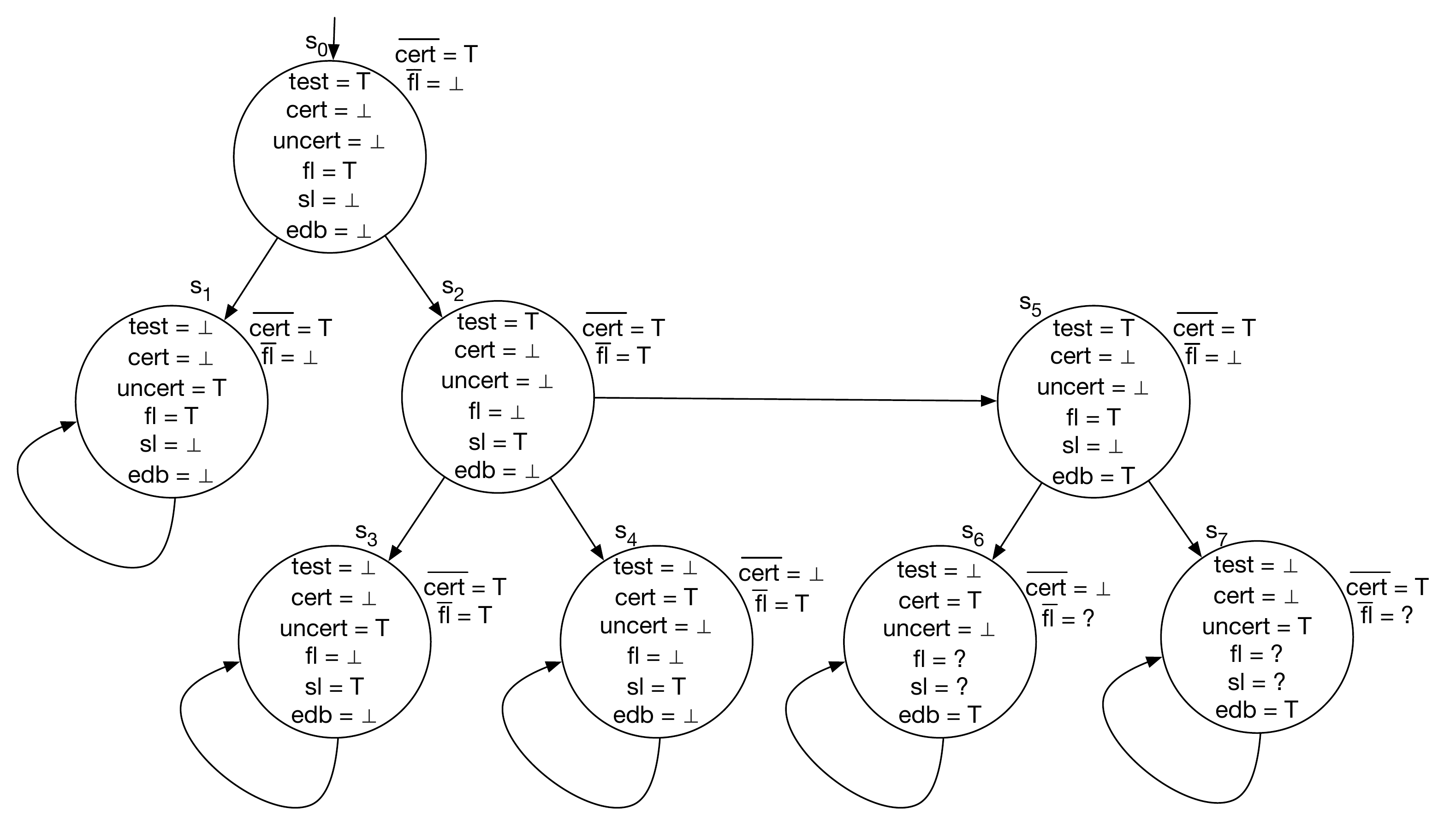}
\end{center}
\caption{Model $M$}  
\label{fig:model}
\end{figure}

We provide in Figure~\ref{fig:model} the complete Partial Kripke Structure that represents the system. 
In each state the propositions are indicated with their truth value.
In the complete version proposed in \cite{arcaini2015formal} all propositions had a true/false value. Note that, in this abstracted version, in the states $s_6$ and $s_7$ the propositions related to the assessed level have an \textit{unknown} value, meaning that the designer is currently not sure on whether the propositions should be $true$ or $false$ in these states.

The propositions $\overline{fl}$ and $\overline{cert}$ that are specified on the side of each state are the complement-closed version of $fl$ and $cert$. These propositions  are used by \NAME\ during the computation of the intersection of the model states with the property states.

\begin{figure}[h!]
\begin{center}         
\includegraphics[width=8.3cm]{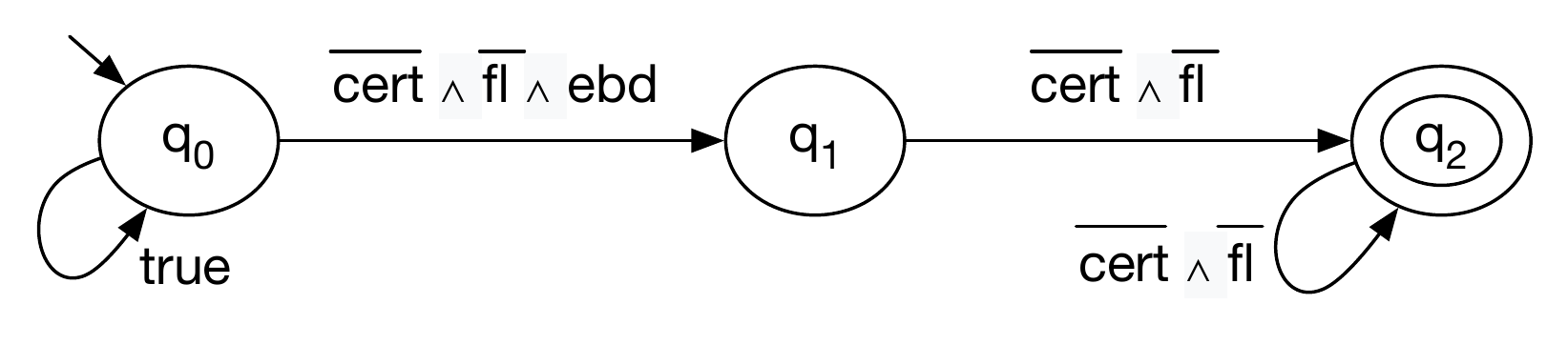}
\end{center}
\caption{The automaton $\mathcal{A}_{\lnot\psi_3}$}  
\label{fig:property}
\end{figure}

\vskip 0.05in  
\textbf{Property.} The property of interest is expressed by the LTL formula $\psi_3=\LTLglobally(edb \rightarrow \LTLfinally (cert \LTLor fl))$, which states that, if an error has been made by the patient (\emph{edb}) he/she cannot be uncertified and be at the second level ($\neg fl$). Indeed, a mistake prevents a patient from increasing the assessed level.
Figure~\ref{fig:property} represents the B{\"u}chi automaton corresponding to $\lnot \psi_3$.

\vskip 0.05in  
\textbf{Running \NAME.}
First, the framework performs a classical model checking run on the pessimistic approximation (generated by assigning $\LTLfalse$ to the propositions $fl$ and $sl$ in the mentioned two states of the model). This particular assignment allows the system to reach the accepting state of the negated property $q_2$ in which holds $\eta(q_2)=\LTLglobally(\lnot cert \LTLand \lnot fl)$. The returned counterexample corresponds to the path $s_0,s_2,s_5,s_7^{\omega}$ on the states of the model, and to the path  $\langle s_0,q_0\rangle,\langle s_2,q_0\rangle,\langle s_5,q_0\rangle,$ $\langle s_7,q_0\rangle,\langle s_7,q_1\rangle,\langle s_7,q_2\rangle^{\omega}$ on the states of the intersection space $M_{pes}\otimes\mathcal{A}_{\lnot\psi_3}$. The generated accepting loop leads to conclude that $M_{pes}\not\models \psi_3$.

\begin{figure}[h!]
\begin{center}         
	\includegraphics[width=9cm]{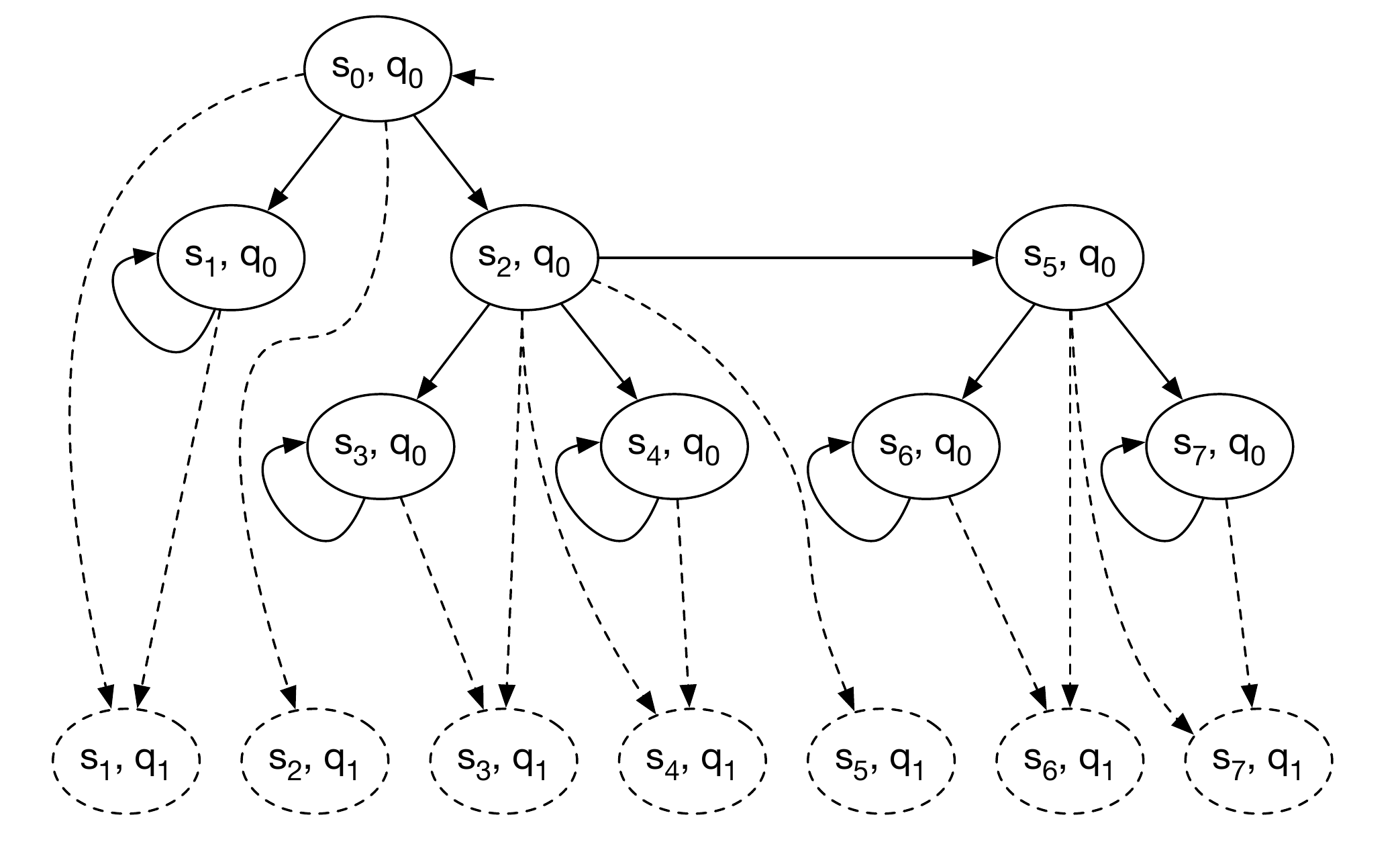}
\end{center}
\caption{The intersection automaton $M_{opt}\otimes\mathcal{A}_{\neg\psi_3}$}  
\label{fig:iopt}
\end{figure}

The framework therefore performs another model checking run on the optimistic approximation (assigning $\LTLtrue$ to the unknown propositions $fl$ and $sl$). This time the intersection state space does not contain any accepting behavior with respect to the negation of property $\psi_3$. 
The intersection space is represented in Figure~\ref{fig:iopt}. Table~\ref{table:proof} presents a formal proof which shows that the optimistic approximation satisfies the  property under analysis.

\begin{longtable}{ | p{2cm} | p{3.5cm} | p{6cm} |  }
\caption{Proof that $\psi_3$ is not violated.}
\label{table:proof}
\endfirsthead
\endhead
\hline
\textbf{Step} &
\textbf{Component} & 
\textbf{Rule} \\
 \hline 
\textbf{Fail}  
&
  $\langle s_1,q_1\rangle,$\newline
  $ \langle s_2,q_1\rangle,$\newline
  $ \langle s_3,q_1\rangle,$\newline
  $ \langle s_4,q_1\rangle,$\newline
  $ \langle s_5,q_1\rangle,$\newline
  $ \langle s_6,q_1\rangle,$\newline
  $ \langle s_7,q_1\rangle$
 &
 \inferrule{
 s_1 \in \mathcal{F}(I_{opt})\\\\
 s_2 \in \mathcal{F}(I_{opt})\\\\
 s_3 \in \mathcal{F}(I_{opt})\\\\
 s_4 \in \mathcal{F}(I_{opt})\\\\
 s_5 \in \mathcal{F}(I_{opt})\\\\
 s_6 \in \mathcal{F}(I_{opt})\\\\
 s_7 \in \mathcal{F}(I_{opt})
}{ s_1\models \mu(q_1)=\lnot edb \LTLor \LTLnext \LTLfinally(cert \LTLor fl) \\\\
 s_2\models \mu(q_1)=\lnot edb \LTLor \LTLnext \LTLfinally(cert \LTLor fl) \\\\
 s_3\models \mu(q_1)=\lnot edb \LTLor \LTLnext \LTLfinally(cert \LTLor fl) \\\\
 s_4\models \mu(q_1)=\lnot edb \LTLor \LTLnext \LTLfinally(cert \LTLor fl) \\\\
 s_5\models \mu(q_1)=\lnot edb \LTLor \LTLnext \LTLfinally(cert \LTLor fl) \\\\
 s_6\models \mu(q_1)=\lnot edb \LTLor \LTLnext \LTLfinally(cert \LTLor fl) \\\\
 s_7\models \mu(q_1)=\lnot edb \LTLor \LTLnext \LTLfinally(cert \LTLor fl)
 }\\	
  \hline 
\textbf{Induction} 
&
 $\mathcal{X}_1=\{\langle s_6,q_0\rangle\},$ \newline
 $Exit(\mathcal{X}_1)=\{\langle s_6,q_1\rangle\}$
  & 
\inferrule{
s_6\models \mu(q_1)  \\\\
s_6\rightarrow \{s_6\}}
 {s_6\models \mu(q_0)=\LTLglobally(edb\rightarrow\LTLfinally(cert\LTLor fl)) } \\
\hline 
 \textbf{Induction} 
 &
 $\mathcal{X}_2=\{\langle s_7,q_0\rangle\},$ \newline
 $Exit(\mathcal{X}_2)=\{\langle s_7,q_1\rangle\}$
  & 
\inferrule{s_7\models \mu(q_1)  \\\\
s_7\rightarrow \{s_7\}}
 {s_7\models \mu(q_0)=\LTLglobally(edb\rightarrow\LTLfinally(cert\LTLor fl)) } \\
\hline 
\textbf{Induction} 
 &
 $\mathcal{X}_3=\{\langle s_3,q_0\rangle\},$ \newline
 $Exit(\mathcal{X}_3)=\{\langle s_3,q_1\rangle\}$
  & 
\inferrule{s_3\models \mu(q_1)  \\\\
s_3\rightarrow \{s_3\}}
 {s_3\models \mu(q_0)=\LTLglobally(edb\rightarrow\LTLfinally(cert\LTLor fl)) } \\
\hline 
\textbf{Induction} 
 &
 $\mathcal{X}_4=\{\langle s_4,q_0\rangle\},$ \newline
 $Exit(\mathcal{X}_4)=\{\langle s_4,q_1\rangle\}$
  & 
\inferrule{s_4\models \mu(q_1)  \\\\
s_4\rightarrow \{s_4\}}
 {s_4\models \mu(q_0)=\LTLglobally(edb\rightarrow\LTLfinally(cert\LTLor fl)) } \\
\hline 
\textbf{Induction} 
 &
 $\mathcal{X}_5=\{\langle s_1,q_0\rangle\},$ \newline
 $Exit(\mathcal{X}_5)=\{\langle s_1,q_1\rangle\}$
  & 
\inferrule{s_1\models \mu(q_1)  \\\\
s_1\rightarrow \{s_1\}}
 {s_1\models \mu(q_0)=\LTLglobally(edb\rightarrow\LTLfinally(cert\LTLor fl)) } \\
\hline 
\textbf{Successors} 
 &
 $\langle s_5,q_0\rangle$
  & 
\inferrule{s_5\rightarrow \{s_6,s_7\}\\\\
s_6\models \mu(q_0) \LTLand  \mu(q_1) \\\\
s_7\models \mu(q_0) \LTLand  \mu(q_1)}
 {s_5\models \mu(q_0)=\LTLglobally(edb\rightarrow\LTLfinally(cert\LTLor fl)) } \\
\hline 
\textbf{Successors} 
 &
 $\langle s_2,q_0\rangle$
  & 
\inferrule{s_2\rightarrow \{s_3,s_4\}\\\\
s_3\models \mu(q_0) \LTLand  \mu(q_1) \\\\
s_4\models \mu(q_0) \LTLand  \mu(q_1)}
 {s_2\models \mu(q_0)=\LTLglobally(edb\rightarrow\LTLfinally(cert\LTLor fl)) } \\
\hline 
\textbf{Successors} 
 &
 $\langle s_0,q_0\rangle$
  & 
\inferrule{s_0\rightarrow \{s_1,s_2\}\\\\
s_1\models \mu(q_0) \LTLand  \mu(q_1) \\\\
s_2\models \mu(q_0) \LTLand  \mu(q_1)}
 {s_0\models \mu(q_0)=\LTLglobally(edb\rightarrow\LTLfinally(cert\LTLor fl)) } \\
\hline 

\underline{Conclusion} 
&
& $s_0\models \mu(q_0) \Rightarrow s_0 \models \psi_3 \Rightarrow M \models \psi_3$
\\\hline
\end{longtable}

The proof can be started by showing first that the system trivially models the property in the states in which $\lnot edb \LTLor \LTLnext \LTLfinally(cert \LTLor fl)$ holds (the fail axiom is applied). Starting from these states, where the model satisfies the property, by using first the induction rule then the successors rule, the proof traverses the automaton until it reaches the initial state. All the premises lead to conclude that it satisfies the $\mu(q_0)$ sub-formula. By construction of the proof \cite{peled2001model}, we can conclude that $s_0$ models the property, i.e., $M_{opt} \models \LTLglobally(edb \rightarrow \LTLfinally (cert \LTLor fl))$.
According to the three-valued model checking algorithm, the result of the procedure is \emph{maybe}, therefore the satisfaction of the property depends on the truth value that will be assigned to the proposition $fl$ in the two uncertain states.

\bibliographystyle{abbrv}
\bibliography{sigproc}


%
%

\end{document}